\documentclass{iau}
\usepackage{graphicx}

\title[Atmospheric Parameters] 
{Atmospheric parameters of individual components of the visual triple stellar system
HIP 32475}

\author[Abdallah M. Hussein et. al]   
{Abdallah M. Hussein$^1$,
 Enas M. Abu-Alrob$^1$, Fatima M. Alkhateri$^2$, \and Mashhoor A. Al-Wardat$^{2,1}$}

\affiliation{$^1$Department of  Physics, Faculty of Sciences, Al al-Bayt University, PO Box: 130040, Mafraq, 25113 Jordan \\[\affilskip]
$^2$ Department of Applied Physics and Astronomy, College of Sciences, and\\ Sharjah Academy for Astronomy, Space Sciences and Technology, University of Sharjah, Sharjah 27272, UAE}

\pubyear{2022}
\volume{FM09}  
\pagerange{119--126}
\jname{IAUGA XXXI}
\editors{A.C. Editor, B.D. Editor \& C.E. Editor, eds.}
\begin{document}

\maketitle

\begin{abstract}
We present a complete analysis of the individual components of the ABC visual triple system HIP 32475. AB pair was discovered during the Hipparcos mission, with a separation of 412 mas. Later, in 2015, a third component was added to the system by discovering it at a  small angular distance from B. In our analysis, we follow Al-Wardat's method for analyzing binary and multiple stellar systems, which is a computational spectrophotometric method. Using estimated parameters, the components' positions on the H-R diagram, evolutionary tracks, and isochrones are defined.  Depending on the analysis,  we estimate the age of the system as 1.259 Gyr with a metallicity of $Z=0.019$. The results show that component A started to evolve from the main sequence to the sub-giants stage, while components B and C are still in the main sequence stage.
\end{abstract}


\section{Introduction}
Recent studies showed that more than half of the stars in the Solar neighborhood are binary or multiple systems (BMSSs) (\cite[{Duquennoy} \& {Mayor}(1991)]{1991A&A...248..485D}; \cite[Raghavan et al. (2010)]{2018AJ....156...31M}; \cite[{Duch{\^e}ne} \& {Kraus} (2013)]{2013ARA&A..51..269D}; \cite[Howell et al. (2021)]{2021FrASS...8...10H}). The architecture of multiple stellar systems and in particular hierarchical is interesting for several reasons, most notably its ability to organize stars into structures that can be interpreted as models of different levels of organization within the universe. Period ratios, mutual orbit orientation, and distribution of masses are all related to the processes of formation and early dynamical evolution of these systems.

This information can tell us a great deal about star formation in general. Most stars are born in groups, so the properties of binary and higher-order systems are very relevant. The life and death of stars are also affected by their companions in various ways. Close pairs of entities can form within hierarchical systems over extended periods of time due to their dynamic evolution (\cite[Tokovinin (2018)]{2018ApJS..235....6T}).\\

Therefore, the systematic examination of multiple stellar systems in order to determine the physical and binary parameters of the components is an essential tool for understanding the origin, evolution, and dynamics of stellar populations as well as the mechanisms of star formation. Many scientific studies demonstrate the importance of this (\cite[Horch et al. (2009)]{2009AJ....137.5057H}; \cite[Tokovinin (2011)]{2011AJ....141...52T}; \cite[{Tokovinin} \& {Horch}(2016)]{2016AJ....152..116T}; \cite[Docobo et al. (2017)]{2017MNRAS.469.1096D}; \cite[Al-Wardat et al. (2021)]{2021RAA....21..161A}).\\

The Gaia and Hipparcos space-astrometric missions of the European Space Agency (ESA) are cornerstones of astrophysical research in various areas of multiple stellar systems, the Milky Way galaxy, unresolved galaxies, exoplanets, stellar evolution, and star-formation theories (\cite[Perryman et al. (1997)]{1997A&A...323L..49P}; \cite[Frankowski et al. (2007)]{2007A&A...464..377F}; \cite[Perryman(2009)]{perryman2009astronomical}; \cite[Gaia Collaboration et al. (2016)]{2016A&A...595A...1G}). These missions provide invaluable information about stars and stellar systems, such as their trigonometric parallax. This information allows for a distance to be determined.\\

In this study, we follow Al-Wardat’s method for analyzing BMSSs to estimate the atmospheric parameters of the individual components of multiple stellar systems HIP 32475 (\cite[Al-Wardat (2002)]{2002BSAO...53...51A}; \cite[Al-Wardat (2012)]{2012PASA...29..523A}). In addition to the estimation of the fundamental parameters of the individual components of BMSSs, Al-Wardat's method provides important information about the accuracy of orbital parameters for multiple systems, as well as testing the accuracy of parallax measurements (\cite[Hussein et al. (2021)]{2022AJ....163..182H}).

\section{Analysis of HIP 32475}

The system HIP 32475 (WDS 06467+0822 = HDS 940AB) is a close visual triple system consisting of the main component A and the  subsystem (BC), in an almost equilateral triangular configuration. The AB pair was discovered during the Hipparcos mission, with a separation of 412 mas. In 2015,  a third component resolved the system by discovering it at a small angular distance from component B. The main data needed to analyze the system are:  the combined (entire) visual magnitude ($m_V=7.04$), which was taken from (\cite[Perryman et al. (1997)]{1997A&A...323L..49P}), the color index ($B-V=0.375\pm0.015$), taken from (\cite[Perryman et al. (1997)]{1997A&A...323L..49P}), Gaia DR3 trigonometric parallax measurement as ($\pi_{G3}=13.7489\pm 0.0974$)  from (\cite[Gaia Collaboration et al. (2021)]{2021A&A...649A...6G}), and the Iron abundance ratio relative to the Sun ($\rm[Fe/H]=0.07$ dex) from
\cite[ G{\'a}sp{\'a}r et al. (2016)]{2016ApJ...826..171G}).

\cite[Tokovinin (2018)]{2018ApJS..235....6T} suggested in the Multiple star catalogue (MSC)  an apparent magnitude and individual mass of component A as [$m_V^{Aa}=7.15$ mag, and $\mathcal{M}_{A} =1.49 \mathcal{M}_{\odot}$],  for the  component Ab as [$m_V^{B}=11.7$ mag, and $\mathcal{M}_{B} =0.72\mathcal{M}_{\odot}$], and for the  component B as [$m_V^{C}=12$ mag, and $\mathcal{M}_{C} =0.71\mathcal{M}_{\odot}$], and the total  mass sum of the system $\Sigma \mathcal{M}$=2.92$\mathcal{M}_{\odot}$. The used magnitude difference for the main system (A,BC) is  $\triangle m_V^{A,BC}= 3.94$ mag, and for the subsystem (Aa,Ab) as $\triangle m_V^{B,C}=0.3$ mag respectively.

\begin{table}
	\centering
	\caption{Magnitudes and colour indices  of the composed synthetic spectrum and individual components for Hip 32475.} \label{sed}
	\resizebox{\linewidth}{!}{%
	\begin{tabular}{cccccccccccccccccc}
	\hline\hline\
	
	&\multicolumn{7}{c}{Johnson}&\multicolumn{7}{c}{Str\"{o}mgren}&\multicolumn{3}{c}{Tycho}
 \\
	
	 &$U$ & $B$&$V$ &$R$&$U-B$&$B-V$& $V-R$   & $u$&$v$ &$b$&$y$  &$u-v$&	$b-y$& $b-y$&$B_T$& $V_T$&$B_T-V_T$
   \\
 
		\hline
\multicolumn{18}{c}{HIP 32475}\\
	\hline
E& 7.42&7.41&7.04&6.82&0.01&0.375&0.22&8.64&7.67&7.26& 7.01&0.97&0.41&0.24&7.50&.09&
     0.41\\
A&7.43&7.43&7.07&6.86&0.0027& 0.36&0.21&8.65&7.68&7.28&7.04&0.97&0.40&
0.23&7.51&7.11&0.39
 \\
B&13.61&12.70&11.61&10.98&0.92&1.08&0.64&14.88&13.34&12.18&11.54&1.54&1.17&0.63&12.99&11.74&1.24 \\
C&14.07&13.06&11.91&11.23&1.01&1.15&0.68&15.39&13.74&12.51&11.84&1.64&1.23&0.67&13.36&12.06&1.30\\
\hline\hline
	\end{tabular}}
	\\
	
\end{table}

Using the calculated input parameters, Al-Wardat’s method builds synthetic spectral energy distributions (SED) for each component alone and for the entire system s a whole  (\cite[Al-Tawalbeh et al. (2021)]{2021AstBu..76...71A}). 
The entire synthetic SED is then used to calculate the synthetic photometry and color indices, which can be compared with observational ones in an iteration process until we reach the best fit. Hence, the input parameters of the individual components represent well their properties. The result parameters are listed in Table.~\ref{tab:phy}.   Depending on these atmospheric and physical parameters, we plot the components   of the  system on the H-R diagram with evolutionary tracks and isochrones (\cite[Masda et al. (2019)]{2019AstBu..74..464M}).

\begin{figure}[ht]
	\centering
	\includegraphics[width=12cm]{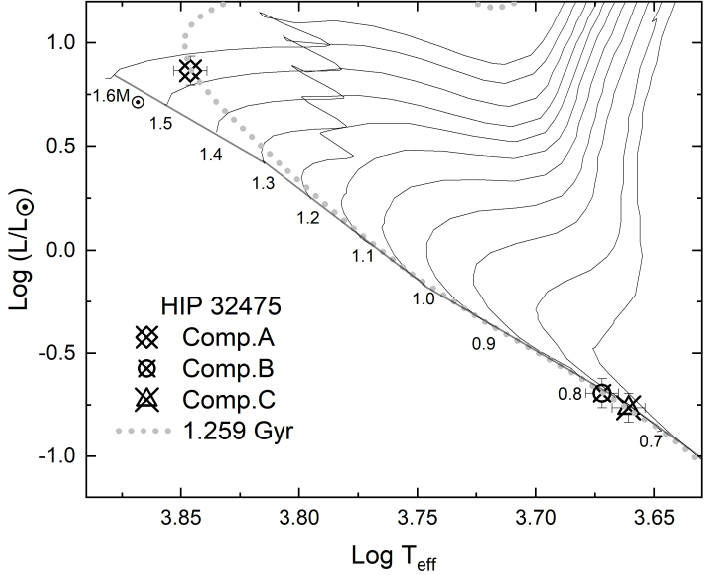}
	\caption{The positions of components of the system on the H-R diagram and evolutionary tracks for solar metallicity. The 1.259 Gyr isochrone is shown in a gray dotted line. } \label{Age}
\end{figure}

Figure.~\ref{Age} shows the position of the system components on the H-R diagram with evolutionary tracks and 1.259 Gyr isochrone. The Figure shows that component A starts to evolve off from the main sequence, while components B and C are still on the main sequence. In addition, the three components are located on the same isochrone (1.259 Gyr). From that, fragmentation is proposed as the most likely process for the formation and evolution of HIP 32475.

\begin{table*}[ht]
\centering
	\small
	\begin{center}
		\caption{The estimated atmospheric parameters of the individual components of HIP 32475.}
		\label{tab:phy}
  	\resizebox{\linewidth}{!}{%
		\begin{tabular}{cccccccccccccc}
			\noalign{\smallskip}
			\hline\hline
			
			Comp.& $ T_{\rm eff}$ [~K~] & $ R$ [R$_{\odot}$]  & $\log g$ [$\rm cm/s^2$] &$ L $ [$\rm L_\odot$] &$ M_{bol}$[mag]&$ M_{V}$[mag] & $\mathcal{M}$[$\mathcal{M}_{\odot}$]&Sp. Type&Age[Gyr] \\
			&{$\pm 80$}&{$\pm 0.09$}&{$\pm 0.07$}&{$\pm0.07$}&{$\pm0.09$}&{$\pm 0.08$}\\
				\hline

\multicolumn{10}{c}{\textbf{HIP 32475}}\\
	\hline
	A&7015&1.835&4.09&7.32&2.59&2.60&1.53&F1V&1.259\\
	B&4700&0.68&4.63&0.20&6.48&6.98&0.76&K3V\\
	C&4580&0.66&4.67&0.17&6.66&7.27&0.74&K4V\\

			\hline\hline
			\noalign{\smallskip}
		\end{tabular}}\\
	
	\end{center}
 \end{table*}

The mass sum from Al-Wardat’s method is given as $\Sigma \mathcal{M}$=3.03$\mathcal{M}_{\odot}$, this result agrees with the mass sum from MSC ($\Sigma \mathcal{M}_{MSC}$=2.91 $\mathcal{M}_{\odot}$) and the dynamical mass calculated using the orbital parameters of 
(\cite[Cvetkovi{\'c} et al. (2020)]{2020AJ....160...48C}) as ($\Sigma \mathcal{M}_{dyn}^{G3}$=3.70$\mathcal{M}_{\odot}$) . Hence, the orbital period ($P=128.949\pm 3.6$ years) and semi-major axis ($a=0.5431\pm 0.0326$) from the orbital solution of the system given by \cite[Cvetkovi{\'c} et al. (2020)]{2020AJ....160...48C}, along with the mass sum from our analysis, we calculate a new  dynamical parallax using Kepler's third law as $\pi_{\rm dyn}=14.7046\pm 0.50$ mas.

\section{Conclusion}

The fundamental parameters of the close visual triple system HIP 32475 are estimated    using Al-Wardat’s method for analyzing BMSSs.
The results confirm the orbit given by \cite[Cvetkovi{\'c} et al. (2020)]{2020AJ....160...48C}. 
The components' positions on the H-R diagram, evolutionary tracks, and isochrones were located, using the estimated parameters.
The results show that component A started to evolve from the main sequence to the subgiant stage, while components B and C are still on the main sequence stage. We estimated the age of the system as 1.259 Gyr. Fragmentation is proposed as the most likely process for the formation and evolution of the system. The estimated masses in this work along with the recent orbital solutions were used to calculate new dynamical parallaxes for the system as $\pi_{\rm dyn}=14.7046\pm 0.50$ mas.

\end{document}